# Ru doping induced spin frustration and enhancement of the room-temperature anomalous Hall effect in La$_{2/3}$Sr$_{1/3}$MnO$_3$ films

*Enda Hua,[#] Liang Si,[#] Kunjie Dai, Qing Wang, Huan Ye, Kuan Liu, Jinfeng Zhang, Jingdi Lu, Kai Chen, Feng Jin, Lingfei Wang,\* and Wenbin Wu\**

#These authors contributed equally to this manuscript

Enda Hua, Kunjie Dai, Qing Wang, Huan Ye, Kuan Liu, Jinfeng Zhang, Jingdi Lu, Feng Jin
Hefei National Research Center for Physical Sciences at Microscale, University of Science and Technology of China, Hefei 230026, China

Liang Si
School of Physics, Northwest University, Xi'an 710127, China
Institut für Festkörperphysik, TU Wien, 1040 Vienna, Austria

Kai Chen
National Synchrotron Radiation Laboratory, University of Science and Technology of China, Hefei 230026, China

Lingfei Wang
Hefei National Research Center for Physical Sciences at Microscale, University of Science and Technology of China, Hefei 230026, China
E-mail: wanglf@ustc.edu.cn

Wenbin Wu
Hefei National Research Center for Physical Sciences at Microscale, University of Science and Technology of China, Hefei 230026, China
Institutes of Physical Science and Information Technology, Anhui University, Hefei 230601, China.
Collaborative Innovation Center of Advanced Microstructures, Nanjing University, Nanjing 210093, China
E-mail: wuwb@ustc.edu.cn

Keywords: transition-metal-oxide, thin film, anomalous Hall effect, spin frustration, skew-scattering

**Abstract**

In transition-metal-oxide heterostructures, the anomalous Hall effect (AHE) is a powerful tool for detecting the magnetic state and revealing intriguing interfacial magnetic orderings. However, achieving a larger AHE at room temperature in oxide heterostructures is still challenging due to the dilemma of mutually strong spin-orbit coupling and magnetic exchange interactions. Here, we exploit the Ru doping-enhanced AHE in $La_{2/3}Sr_{1/3}Mn_{1-x}Ru_xO_3$ epitaxial films. As the B-site Ru doping level increases up to 20%, the anomalous Hall resistivity at room temperature can be enhanced from $n\Omega \cdot cm$ to $\mu\Omega \cdot cm$ scale. Ru doping leads to strong competition between ferromagnetic double-exchange interaction and antiferromagnetic super-exchange interaction. The resultant spin frustration and spin-glass state facilitate a strong skew-scattering process, thus significantly enhancing the extrinsic AHE. Our findings could pave a feasible approach for boosting the controllability and reliability of oxide-based spintronic devices.

# 1. Introduction

Anomalous Hall effect (AHE), originating from the interplay between spin-orbit coupling (SOC) and time-reversal symmetry breaking, is one of the most fundamental and intriguing transport phenomena in magnetically-ordered systems.[1,2] Over the past decades, AHE has been discovered and intensively investigated in various ferromagnets,[3,4] noncollinear antiferromagnets,[5-7] and even nonmagnetic system.[8,9] There are currently three main mechanisms that contribute to the AHE, the intrinsic, skew-scattering, and side-jump contributions.[1] The intrinsic contribution, proposed by Karplus and Luttinger,[10] considers that electrons gain an anomalous velocity perpendicular to the external electric field, and the magnitude is solely determined by the integral of the $k$-space Berry curvature over the Brillouin zone. The skew-scattering mechanism considers an asymmetric scattering of spin-up and spin-down electrons due to the effective SOC of impurities or electrons.[11] The side-jump mechanism considers that the spin-up and spin-down electrons can be deflected in opposite directions as they approach and leave spin-orbit coupled impurities.[12] The skew-scattering and side-jump are classified as the extrinsic mechanisms for AHE. In a real magnetic system, these intrinsic and extrinsic mechanisms usually coexist and cooperatively contribute to the experimentally observed AHE signals.

For transition-metal-oxide (TMO) -based heterostructures, the AHE is particularly essential for both fundamental research and device applications. On one hand, for these strongly-correlated interface systems, the AHE is an ideal fingerprint that manifests the complex interplay between charge, spin, orbit, and lattice degrees of freedom. For instance, the significant AHE signals observed in the 3d/5d oxide heterostructures prove the existence of interfacial charge transfer and ferromagnetic proximity effect.[13-15] In the $SrRuO_3$,[16-19] $SrCoO_3$,[20] and $EuTiO_3$[21] films, the nonmonotonic magnetic field and temperature-dependent AHE not only signify the symmetry-protected nodal points in the band structure but also imply the spatial inhomogeneities in magnetism. On the other hand, AHE is an essential tool to read out the magnetic state in oxide-based spintronic devices. In the $SrIrO_3$ and $SrRuO_3$-based spin-transfer and spin-orbit torque devices, the current-induced magnetization switching can be identified by the sharp changes in anomalous Hall voltages.[22-24]

Despite the essential roles of the AHE, it is still challenging to achieve a large AHE at room temperature in TMO systems. For the 4d or 5d TMO-based heterostructures (e.g., ruthenates or iridates), the strong SOC leads to large anomalous Hall resistivity ($\rho_{AHE}$) up to

μΩ·cm, but the weak exchange coupling strength limits the Curie temperature ($T_C$) below room temperature by more than 100 K.[13,17,18] In contrast, the 3d TMO (i.e. manganites or ferrites) has strong exchange coupling strength and thus high $T_C$ well above room temperature.[25,26] However, the weak SOC limited $\rho_{AHE}$ on the scale of nΩ·cm, which inevitably hinders the electrical detection of magnetic states in practical spintronic devices.

In this work, to circumvent the contradiction between high $T_C$ and large AHE, we exploit a route to enhance the AHE of La$_{2/3}$Sr$_{1/3}$MnO$_3$ (LSMO) epitaxial thin films by Ru doping. By replacing 20% of the B-site Mn with Ru, the saturated $\rho_{AHE}$ at room temperature can be enlarged up to the μΩ·cm scale. The enhanced AHE is dominated by the extrinsic skew-scattering process rather than the intrinsic Berry phase contribution. We further investigated the Ru doping effects on the electronic and magnetic structures at the microscale. The doping-induced competition between Ru-Mn antiferromagnetic interaction and Mn-Mn ferromagnetic interaction could cause considerable spin frustration. The resultant spin-glass state, together with the Ru scattering centers with strong SOC, could be the main driving force of strong skew-scattering and enhanced AHE. Our work paved an experimental route to enhance the room-temperature AHE in TMO-based ferromagnets, which facilitates effective read-out and harnessing of the magnetic states in oxide-based spintronic devices.

## 2. Results and Discussions
### 2.1. Film growth and structural characterization

La$_{2/3}$Sr$_{1/3}$MnO$_3$ (LSMO) and La$_{2/3}$Sr$_{1/3}$Mn$_{1-x}$Ru$_x$O$_3$ (LSMRO) films are epitaxially grown on the (001)-oriented (LaAlO$_3$)$_{0.3}$(SrAl$_{0.5}$Ta$_{0.5}$O$_3$)$_{0.7}$ [LSAT(001)] substrates by pulsed laser deposition. The Ru dopant can replace the B-site Mn in the LSMRO. We fix the film thickness at 30 nm and limit the range of Ru doping level ($x$) from 0 to 0.20. Further increasing Ru doping will cause considerable disorder in the film structure and degradation in the metallicity, which hinders our study on the AHE.[27] The details about film growth can be found in the Method Section. Note that LSMRO compounds for the entire doping range ($x$ = 0 ~ 0.20) have rhombohedral unit-cells, while the LSAT(001) substrate has a cubic unit-cell. For clarity, we utilize pseudocubic notations for both substrates and films in the following parts of this paper. The pseudocubic lattice constants of LSAT ($a_s$) and bulk LSMRO ($a_b$) for various $x$ are summarized in Table 1. According to the film/substrate lattice mismatches [$\varepsilon_f = (a_b - a_f) / a_b$], the LSAT(001) substrates should impose compressive strain on all of the LSMRO films.

| Bulk materials | $a_c$ or $a_b$ (Å) | $\varepsilon_f$ (%) |
|---|---|---|
| LSAT | 3.868 | |
| La$_{2/3}$Sr$_{1/3}$MnO$_3$ | 3.874 | -0.155 |
| La$_{2/3}$Sr$_{1/3}$Mn$_{0.95}$Ru$_{0.05}$O$_3$ | 3.877 | -0.233 |
| La$_{2/3}$Sr$_{1/3}$Mn$_{0.90}$Ru$_{0.10}$O$_3$ | 3.884 | -0.414 |
| La$_{2/3}$Sr$_{1/3}$Mn$_{0.85}$Ru$_{0.15}$O$_3$ | 3.886 | -0.465 |
| La$_{2/3}$Sr$_{1/3}$Mn$_{0.80}$Ru$_{0.20}$O$_3$ | 3.893 | -0.646 |

**Table 1**. Pseudocubic lattice constants of LSAT ($a_s$) and bulk LSMRO ($a_b$) for all the doping levels $x$. The lattice mismatches between the substrate and films [$\varepsilon_f = (a_b - a_f)/a_b$] are also calculated. The negative values imply a compressive strain state.

We first examine the epitaxial qualities of LSMRO/LSAT(001) films by X-ray diffraction (XRD) and atomic force microscopy (see Figure S1 of the Supporting Information). The XRD *2θ-ω* linear scans of 30 nm LSMRO films with various $x$ display well-defined Laue fringes around the LSMRO(002) diffractions, signifying sharp LSMRO/LSAT interfaces. The full width at half maximum (FWHM) of the *ω*-scan rocking curves near LSMRO(002) diffractions for all of the samples are below 0.02°. The surface topography image shows a well-organized one-unit-cell-high terrace structure. These results further confirm the exceptionally high epitaxial qualities of our LSMRO films. The off-specular reciprocal space mappings (Figure S2 of the Supporting Information) show that the in-plane reciprocal space vectors for all of the LSMRO films are the same as those of LSAT(001) substrates, demonstrating the coherent strain states. Moreover, as the Ru doping level $x$ increases, the LSMRO(002) peaks gradually shift toward lower Bragg angles, signifying a lattice elongation along the out-of-plane *c*-axis. These structural modulations can be attributed to both the compressive strain and the larger ion radius of Ru$^{4+}$/Ru$^{3+}$ compared to Mn$^{4+}$/Mn$^{3+}$.

## 2.2 Electrical transport and magnetic properties

Ru doping can significantly modulate the electrical transport and magnetic properties of LSMO films. **Figure 1**a and 1b show the temperature-dependent resistivity (*ρ-T*) and magnetization (*M-T*) curves of 30 nm LSMRO with different $x$. For the films with $x \leq 0.15$, the *ρ-T* and *M-T* curves clearly show paramagnetic insulator-to-ferromagnetic (FM) metal transition upon cooling. As $x$ increases, the $T_C$, electrical conductivity, and saturated magnetization ($M_S$) decrease, suggesting a gradually suppressed FM metal phase. Further increasing $x$ up to 0.2, the additional *ρ* upturn below 100 K as well as the low $M_S$ of ~2.5 μB/Mn indicate that the FM metal phase is more severely suppressed. This behavior has been observed in bulk LSMRO and can be attributed to the competition between Mn-O-Mn double-exchange (DE) and Ru-O-Mn super-exchange (SE) interactions.[27,28]

We also measured the magnetic field-dependent magnetization (*M-H*) curves from the 30 nm LSMRO films at 10 K with *H* applied along the film plane or the film normal (denoted as $H_{//}$ and $H_\perp$, respectively). The *M-H* loops (see Figure 1c) of the LSMO film demonstrate a typical easy-plane magnetic anisotropy (EMA) from the demagnetization effect. In contrast, as shown in Figure 1d, the *M-H* loops of LSMRO (*x* = 0.15) film signify a perpendicular magnetic anisotropy (PMA). Such a PMA becomes stronger as *x* increases (see Figure S3 of the Supporting Information). According to previous works, inducing such a strong PMA in LSMO requires a large compressive strain over 2% [imposed by LaAlO$_3$(001) substrate],[29] while the compressive strain in the LSMRO/LSAT(001) film (*x* = 0.15) is less than 0.5% (Table 1). On this basis, we speculate that Ru doping can enhance the SOC and thus magnify the strain-mediated magnetocrystalline anisotropy of the LSMRO films.

## 2.3 Ru doping-enhanced anomalous Hall effect

Now we turn to characterize the central physical property in this work, namely, the AHE of LSMRO films. **Figure 2**a shows the *H*-dependent anomalous Hall resistivity ($\rho_{AHE}$-*H*) curves of 30 nm LSMRO films at 10 K. The large PMA leads to sharp magnetic switching in the $H_\perp$ case, which helps us to precisely determine the magnitude of saturated $\rho_{AHE}$. The saturated $\rho_{AHE}$ for LSMO film at 10 K is ~ 2 nΩ·cm, which is barely detectable. After 5% Ru doping, surprisingly, the $\rho_{AHE}$ is enhanced to 24 nΩ·cm, more than one order of magnitude higher than that of the LSMO film. Together with the enlarged coercive field, $\rho_{AHE}$ increases continuously with the Ru doping level *x*. For the LSMRO film with *x* = 0.20, the saturated $\rho_{AHE}$ at 10 K reaches ~0.5 μΩ·cm. The $\rho_{AHE}$-*H* curves of the LSMRO (*x* = 0.15) at various *T* are plotted in Figure 2b. Upon warming, the saturated $\rho_{AHE}$ increases and reaches the maximum value at 310 K, which is close to the *T*$_C$. As *T* increases further, the $\rho_{AHE}$-*H* curve becomes much more slanted and the saturated $\rho_{AHE}$ decreases sharply, implying degraded ferromagnetism. Similar trends can be observed in LSMRO films with other *x* values. For the *x* = 0.2 (*x* = 0.15) samples, the $\rho_{AHE}$ reaches the maximum value of 1.23 (0.61) μΩ·cm at 280 K (310 K). These room-temperature $\rho_{AHE}$ values are comparable with those of prototypical 4d ferromagnetic oxide SrRuO$_3$ and can be easily read out in oxide-based spintronic devices.[17] Note that we also observed similar AHE enhancement in the tensile-strained LSMRO/SrTiO$_3$(001) films (Figure S4 and S5 in the Supporting Information). Hence, the effective AHE enhancement in LSMRO films should dominated by the Ru doping effect rather than possible variations of epitaxial strain or MA.

The transverse $\rho_{AHE}$ usually shows a strong correlation with the longitudinal resistivity $\rho_{xx}$. According to the simplified phenomenological scaling analyses, $\rho_{AHE}$ can be expressed by the following equation:[30,31]

$$\rho_{AHE} = \kappa_{sk}\rho_{xx} + (\kappa_{int} + \kappa_{sj})\rho_{xx}^2 \tag{1}$$

The first linear term is the AHE contribution from the skew-scattering mechanism, while the second quadratic term is the AHE contributions from the intrinsic Berry phase and side-jump mechanisms. And $\kappa_{sk}$, $\kappa_{int}$, and $\kappa_{sj}$ are scaling coefficients for the skew-scattering, intrinsic and side-jump terms, respectively. Accordingly, the skew-scattering contribution to the AHE can be identified or separated by analyzing the algebraic power dependences between $\rho_{xx}$ and $\rho_{AHE}$, whereas the intrinsic and side-jump contributions can be separated only if introducing the additional scaling parameters, e.g., the residual resistivity.[32,33] **Figure 3**a-d shows the log-plot of $\rho_{xx}$-$\rho_{AHE}$ curves from 30 nm LSMRO films with various $x$. In the low $T$ region (corresponding to small $\rho_{xx}$ and nearly saturated $M$), all of the $\rho_{xx}$-$\rho_{AHE}$ log-plots are linear. Accordingly, the $\rho_{xx}$-$\rho_{AHE}$ relationship can be simply described by

$$log(\rho_{AHE}) = \alpha log(\rho_{xx}) \tag{2}$$

where $\alpha$ is the slope value obtained from linear fitting of the $\rho_{xx}$-$\rho_{AHE}$ log-plots at the low $T$ region. This equation can be converted to

$$\rho_{AHE} = \rho_{xx}^\alpha \tag{3}$$

The Ru doping level $x$-dependent $\alpha$ value is plotted in Figure 3e. For the LSMO film ($x = 0$), $\alpha \sim 1.8$. The nearly quadratic relationship between $\rho_{AHE}$ and $\rho_{xx}$ is consistent with previous publications, which suggest that the Berry-phase-mediated intrinsic mechanism should dominate the AHE in LSMO.[34,35] As $x$ increases up to 0.15 and 0.20, $\alpha$ decreases gradually and approaches 1.07 and 1.01, respectively. The linear relationship between $\rho_{AHE}$ and $\rho_{xx}$ implies that the dominated AHE mechanism converts from the intrinsic one to the skew-scattering. We further calculated the anomalous Hall angle $\theta_H = \rho_{AHE} / \rho_{xx}$ at 10 K. As shown in Figure 3f, for the samples with $0 \leq x \leq 0.15$, the Ru doping-induced increment of $\theta_H$ is consistent with the reduction of $\alpha$. The slight decrease of $\theta_H$ at $x = 0.20$ could be related to the upturn of $\rho_{xx}$ below 100 K (see Figure 1a). These results further suggest that the Ru doping-enhanced skew-scattering should be the main driving force of the strong AHE observed in LSMRO.

**2.4 Electronic and magnetic structures**

Before exploiting the physical origin of the Ru doping-enhanced AHE, we analyzed the electronic structures of LSMRO films through X-ray absorption spectroscopy (XAS). The XAS curves near the O-K and Mn-L edges are measured from the 30 nm LSMRO films with $x = 0$, 0.05, and 0.10. As shown in Figure S6 of the Supporting Information, the XAS curves at the O-K edge barely changed with $x$, which means that Ru doping should not cause detectable changes in the oxygen content, i.e., introducing additional oxygen vacancies in the LSMRO films. In contrast, both the Mn $L_2$ and $L_3$ peaks exhibit a slight shift to lower energy as $x$ increases. Note that the XAS peaks near the Mn-L edge are strongly correlated with the valence state of Mn cations. Such peak shifts imply that the Ru dopants can donate electrons to the Mn sites.[36] In the parent compound LSMO, the Sr doping level is already optimal for the highest $T_C$. Therefore, Ru doping-induced electron donation is expected to degrade the ferromagnetism, which is consistent with our $\rho$-$T$ and $M$-$T$ curves shown in Figure 1.

We then investigate the Ru doping effect on the electronic and magnetic structure of LSMRO through first-principles calculations based on the density-functional theory plus on-site Coulomb interaction (DFT+$U$) framework. As schematically shown in **Figure 4**a, we construct 2×2×2 LSMO and LSMRO supercells for the calculations (the procedural details about supercell construction and calculation can be found in the Method section). In the LSMRO supercell, the central MnO$_6$ is replaced by a RuO$_6$ octahedron, resulting in a nominal $x = 1/8$. And each RuO$_6$ octahedron is surrounded and isolated by six MnO$_6$ octahedra, close to our experimental case. Further approaching the experimental concentration requires a larger supercell that challenges DFT computational ability. As marked in Figure 4a and Figure S7 of the Supporting Information, we define the first, second, and third nearest Mn sites to the central Mn or Ru as Mn-1, Mn-2, and Mn-3, respectively.

The spin-polarized density of states (DOS) profiles of the Mn $3d$ orbitals in LSMO are shown in Figure 4b. The DOS near the Fermi surface ($E_F$) is dominated by the Mn $e_g$ band. All of the ~3.67 electrons occupy the spin-up (majority) channel while the spin-down (minority) channel is empty, clearly signifying an FM half-metallic behavior. For the LSMRO case, the DOS near the Fermi surface ($E_F$) is dominated by both the Mn $e_g$ and Ru $t_{2g}$ bands. The DOS profiles of Mn $3d$ bands vary slightly for different Mn sites (see Figure S7 of the Supporting Information). Even for the Mn-1 site (see Figure 4c), the DOS profile is rather similar to that in LSMO, which means the Ru doping effect on the LSMRO electronic structure is rather localized. To estimate the possible charge transfer between Ru and Mn cations, we calculated the electron occupation at Mn and Ru sites in the LSMRO supercell. The charges at Mn-1, Mn-2, Mn-3, and Ru sites are 3.77$e$, 3.63$e$, 3.72$e$, and 4.56$e$, respectively. A detectable charge transfer of ~0.11e occurs from Ru to Mn-1, which is consistent with the XAS results. The DOS profiles of Ru $4d$ orbitals further show that the spin-down channel of the Ru $t_{2g}$ band is fully occupied, while the spin-up channel is partially occupied by ~1.56$e$. Thus, charge transfer should occur from the spin-up channel of the Ru $t_{2g}$ band.

In addition to the electronic structures, the DFT+U calculations further suggest that Ru doping can also change the magnetic ground state. For both the LSMO and LSMRO supercells, we calculate the total energies of a variety of magnetic orderings (see Figure 4e,f and Figure S8 of the Supporting Information). For the LSMO case, all kinds of AFM orderings elevate the total energy, demonstrating a robust FM ground state. For the LSMRO case, although the long-range AFM orderings are still unstable, the energy differences between AFM and FM orderings are reduced. More surprisingly, by aligning the magnetic moment of the central Ru site antiparallel to those of the adjacent Mn sites (denoted as AFM-Ru), the total energy can be lowered by ~21 meV/u.c. We further calculate the first, second, and third nearest Mn-Mn (for LSMO) or Mn-Ru (for LSMRO) exchange interactions (defined as $J_1$, $J_2$, and $J_3$, respectively) in both LSMO and LSMRO supercells. As listed in Table 2, the FM ground state of LSMO is stabilized by the negative $J_1$ (-73.4 meV), which overwhelms the positive $J_2$ (+3.2 meV) and $J_3$ (+2.8 meV). By contrast, $J_1$ in LSMRO (+50.1 meV) turns out to be positive. The $J_1$ in LSMRO (between Ru and Mn-1) is smaller than that in LSMO, probably due to the smaller local moment of Ru site (below 2 $\mu_B$, far below the Mn site moment of ~3.67 $\mu_B$). Moreover, unlike the negligible $J_2$ and $J_3$ in LSMO, the negative $J_2$ (-20.4 meV) and positive $J_3$ (+3.6 meV) in LSMRO are much stronger, which may be related to the spatially-extended Ru $4d$ orbitals and larger orbital overlaps (See Table S1 in the Supporting Information for computational details).[17] The AFM $J_1$ (between Ru and Mn-1) and $J_3$ (between Ru and Mn-3) could compete with the FM $J_2$ (between Ru and Mn-2), thus leading to the local AFM ordering (i.e., AFM-Ru ordering).

|  | $J_1$ | $J_2$ | $J_3$ |
|---|---|---|---|
| La$_{2/3}$Sr$_{1/3}$MnO$_3$ | -73.4 meV | +3.2 meV | +2.8 meV |
| La$_{2/3}$Sr$_{1/3}$Mn$_{7/8}$Ru$_{1/8}$O$_3$ | +50.1 meV | -20.4 meV | +3.6 meV |

**Table 2.** DFT+$U$ calculated exchange interactions for LSMO and LSMRO supercells. The first, second, and third nearest Mn-Mn (Mn-Ru) exchange interactions in LSMO (LSMRO) are defined as $J_1$, $J_2$, and $J_3$, respectively. Positive (negative) values indicate the AFM (FM) interaction.

The local AFM coupling between Ru and Mn-1 sites can be understood by the competition between DE and SE interactions. As schematically depicted in Figure 4g, the FM ordering in LSMO is driven by DE interaction. As a spin-up itinerant electron hops from the O site to the $Mn^{4+}$ site, the vacant O $2p$ orbital is then filled by a spin-up $e_g$ electron from the nearest $Mn^{3+}$. The local $t_{2g}$ spins on Mn sites can be aligned parallel according to Hund's rule.[37] For the LSMRO case, on one hand, the high energy Ru $e_g$ levels make the spin-up electron hopping from O $2p$ to $Ru^{4+}$ energetically unfavorable. On the other hand, a hopping for spin-up electrons from O $2p$ to Ru $t_{2g}$ orbitals can be facilitated as the local spins in the Ru minority $t_{2g}$ band aligns anti-parallel to those in Mn $t_{2g}$ orbitals, leading to an AFM SE coupling.

## 2.5 Ru doping-induced spin frustration

To build the connection between local Mn-Ru AFM coupling and enhanced AHE, we further characterized the *M-T* curves after zero-field cooling (ZFC) and field cooling (FC) of the 30 nm LSMRO films with various *x*. For the LSMO film, the *M-T* curves measured after FC and ZFC nearly coincide with each other, consistent with its robust ferromagnetism. For all of the LSMRO films ($x \geq 0.05$), on the contrary, the FC and ZFC curves exhibit obvious bifurcations below the irreversibility temperature ($T_{irr}$). This bifurcation behavior is usually regarded as the signature of spin-glass or cluster glass states.[38-40] As *x* increases, the bifurcation becomes more significant, and the $T_{irr}$ shifts toward higher temperatures (see **Figure 5**a-e). Especially for the $x = 0.15$ and 0.20 cases, the *M* values at low *T* in the ZFC curve are close to 0, indicating a highly disordered or frustrated spin configuration. Figure 5f shows the ZFC and FC *M-T* curves of the 30 nm LSMRO ($x = 0.10$) films measured at various external *H*. As *H* increases, the $T_{irr}$ decreases monotonically. As shown in Figure 5g, the *H*-dependence of $T_{irr}$ follows the Almeida-Thouless (AT) equation:[38]

$$H(T_{irr})/\Delta J \propto (1 - \frac{T_{irr}}{T_F})^{3/2} \tag{4}$$

where $T_F$ is the zero-field spin-glass freezing temperature and $\Delta J$ is the width of the distribution of the exchange interaction. The fitting and AC susceptibility results (see Figure S10 of the Supporting Information) further demonstrate that Ru doping can cause a spin-glass or cluster glass state. Since the $T_{irr}$ and peak temperature in the ZFC *M-T* curves are nearly inseparable in our LSMRO films, we suggest that the Ru doping-induced spin-glass state should be the most feasible scenario.[41] This is also consistent with the unfavorably high energy cost in forming the cluster-like AFM structure (see Figure 4f and Figure S8 of the Supporting Information).

Now, we discuss the physical connections between the spin-glass state and Ru doping-enhanced AHE. As aforementioned in the first-principles calculations on LSMRO, the positive $J_1$ (between Ru and Mn-1) and negative $J_2$ (between Ru and Mn-2) share comparable magnitudes with the negative Mn-Mn $J_1$. In the practical LSMRO films, near the randomly distributed Ru dopants, the competitions among these AFM and FM exchange interactions could intensively compete with each other, probably leading to considerable spin frustrations and the spin-glass state.[39,42,43] Note that similar spin glass behavior are also observed in the LSMRO/SrTiO$_3$(001) films, further highlighting the dominating role of Ru-doping (Figure S11 in the Supporting Information). Compared to the pure FM ordering, the spin-glass state could considerably enlarge the probability of spin-dependent scattering for itinerant electrons.[44-46] And the 4d Ru cations may act as scattering centers with strong SOC. Both scenarios are expected to enhance the skew-scattering and thus the AHE. Moreover, as the Ru doping level $x$ increases, the spin frustration should be promoted, which is consistent with the gradually enhanced AHE (see Figure 2 and 3). In addition, the spin-glass state may also cause some spin chirality and topological contributions to the AHE,[47,48] which is implied by the detectable inconsistency in line shape between the *M-H* and $\rho_{AHE}$-*H* curves.[49] (see Figure S12 of the Supporting Information)

## 3. Conclusion

In summary, we systematically explored the Ru doping enhanced AHE in LSMRO/LAST(001) films. Defying the dilemma of mutually large AHE and high Curie temperature for transition metal oxides, the LSMRO films achieved both. The maximum $\rho_{AHE}$ at room temperature is up to ~1 μΩ·cm, three orders of magnitude higher than that of the parent LSMO compound. Such significant AHE enhancement can be attributed to the Ru doping-induced spin frustration and thus the promoted skew-scattering. Our results provide an experimental route to circumvent the contradiction between the large SOC and the strong magnetic exchange interaction in TMO-based thin-film ferromagnets. Specifically, the high $T_C$ can be ensured by choosing 3d TMO-based ferromagnets and the AHE can be magnified by introducing a certain amount of 4d or 5d transition-metal dopants. This experimental approach is expected to be applicable for various TMO systems and could boost the read-out reliability and manipulation accuracy of the magnetic states in oxide-based spintronic devices.

## 4. Experimental Section

*Synthesis of polycrystalline ceramic target and single-phase films*: The ceramic targets La$_{2/3}$Sr$_{1/3}$Mn$_{1-x}$Ru$_x$O$_3$ (LSMRO) are synthesized by the solid-state reactions. La$_2$O$_3$, SrCO$_3$, MnO$_2$, and Ru$_2$O$_3$ powders were mixed according to the stoichiometry and then calcined sequentially at 1050 and 1250 °C for 12 hours. After pressing into a pellet of 1 inch in diameter at 30 MPa, the targets are sintered at 1350 °C for 24 hours. We grow the LSMRO films with various doping levels ($x$) and film thicknesses by the corresponding ceramic targets on (001)-oriented (LaAlO$_3$)$_{0.3}$(SrAl$_{0.5}$Ta$_{0.5}$O$_3$)$_{0.7}$ [LSAT(001)] substrates by pulsed laser deposition (KrF excimer laser, λ=248 nm). During the film growth, we kept the substrate temperature and oxygen pressure at 680 °C and 45 Pa, respectively. After deposition, we annealed the films in-situ under the growth condition for 15 minutes and then cooled them down to room temperature in an oxygen atmosphere of 2000 Pa.

*Structural, magnetic, and electrical characterizations*: The epitaxial quality of the samples was examined by high-resolution X-ray diffraction (XRD, PANalytical Empyrean X-ray diffractometer, Cu $K_{\alpha 1}$ radiation) with both the 2$\theta$-$\omega$ line scan and off-specular reciprocal space mapping (RSM) mode. The magnetizations were measured on a vibrating sample magnetometer (VSM-SQUID, Quantum Design). The longitudinal and transverse resistivities were characterized by a Physical Properties Measurement System (PPMS, Quantum Design). Before measurements, the films were patterned into 6-probe Hall bars through photolithography. The ordinary Hall resistivity ($\rho_{\text{OHE}}$) was subtracted from the total Hall resistivity ($\rho_{xy}$) by linear fitting the $\rho_{xy}$-$H$ curve in the range of $H > 1$ T.

*X-ray absorption experiments*: Soft X-ray absorption spectroscopy (XAS) was conducted at the Beamlines MCD-A and MCD-B (Soochow Beamline for Energy Materials) in Hefei Light Source (National Synchrotron Radiation Laboratory, NSRL). The X-ray absorption spectra (XAS) of the O-K edge and Mn-L$_{2,3}$ edge were recorded at 300 K in the total electron yield (TEY) mode. All of the spectra were acquired with incident light along the film normal.

*First-principles calculations*: To evaluate the Ru doping effects on the structural, magnetic, and electronic properties of LSMO, we perform first-principles calculations based on the density functional theory + on-site Coulomb interaction (DFT+$U$) frameworks.[50] We first construct a 2×2×2 supercell of LaMnO$_3$. Orthorhombic a$^-$a$^-$c$^+$ octahedral rotation/tilting is set to be the initial structure. A-site Sr doping in the LSMO compound is achieved by employing Virtual crystal approximation (VCA).[51] The Ru doping is achieved by replacing the central Mn in the supercell with Ru, leading to the chemical formula La$_{2/3}$Sr$_{1/3}$Mn$_{7/8}$Ru$_{1/8}$O$_3$ (LSMRO, x = 1/8). Further approaching the experimental concentration requires a much larger supercell, which challenges DFT computational ability. Structural relaxations, magnetic total energy, and density of states (DOS) calculations are carried out through VASP[52] code within PBE[53] versions of the generalized gradient approximation (GGA) on a 7×7×7 k-meth. Wien2k code[54] within PBE is employed for benchmarks. The simplified (rotationally invariant) approach to the DFT+$U$ induced by Dudarev is employed,[55] and the interaction parameters for Mn-d and Ru-d are adopted from our previous studies for 3d nickelates[56] and 4d ruthenates:[57] on-site Coulomb $U$ for Mn and Ru are 4.4 and 3.0 eV, respectively. These values were obtained from our constrained random phase approximation (cRPA)[58] calculations and have been carefully tested for validity.

Before computing the DOS and magnetic ground state, we perform structural relaxations for both LSMO and LSMRO supercells, in which the in-plane lattice constants are fixed to that of LSAT(001) substrate. The differences of lattice constants between the experimental and calculation values are all within 0.5%. The relaxed structures are plotted in Figure 4a and Figure S9 of the Supporting Information (with detailed atomic arrangements). To determine the magnetic ground state, we calculated the total energy of various magnetic orderings, including the ferromagnetic, A-, C-, and G-type antiferromagnetic orderings (denoted as FM, A-AFM, C-AFM, and G-AFM, respectively). In addition to these highly symmetrical magnetic orderings, we also considered the stability of several possible orderings with lower symmetries, including the AFM alignment of the Ru moment only (AFM-Ru), a single-column C-type-like AFM (AFM-C'), and a cluster-like AFM (AFM-cluster). The schematic magnetic structures are plotted in Figure 4e, Figure S8, and S9 of the Supporting Information. Other possible magnetic orderings lead to energetic nonconvergence in the DFT+$U$ calculations. And the effective Mn-Mn and Mn-Ru exchange interactions ($J_1$, $J_2$, and $J_3$) are obtained by solving the total energy-magnetic orderings linear equation systems.

**References**

[1]  N. Nagaosa, J. Sinova, S. Onoda, A. H. MacDonald, N. P. Ong, *Rev. Mod. Phys.* **2010**, 82, 1539.
[2]  S. Onoda, N. Sugimoto, N. Nagaosa, *Phys. Rev. B.* **2008**, 77, 165103.
[3]  S. Onoda, N. Sugimoto, N. Nagaosa, *Phys. Rev. Lett.* **2006**, 97, 126602.
[4]  T. Jungwirth, Q. Niu, A. H. MacDonald, *Phys. Rev. Lett.* **2002**, 88, 207208.
[5]  H. Chen, Q. Niu, A. H. MacDonald, *Phys. Rev. Lett.* **2014**, 112, 017205.
[6]  S. Nakatsuji, N. Kiyohara, T. Higo, *Nature* **2015**, 527, 212.
[7]  L. Šmejkal, A. H. MacDonald, J. Sinova, S. Nakatsuji, T. Jungwirth, *Nat. Rev. Mater.* **2022**, 7, 482.
[8]  T. Liang, J. Lin, Q. Gibson, S. Kushwaha, M. Liu, W. Wang, H. Xiong, J. A. Sobota, M. Hashimoto, P. S. Kirchmann, Z.-X. Shen, R. J. Cava, N. P. Ong, *Nat. Phys.* **2018**, 14, 451.
[9]  K. Kang, T. Li, E. Sohn, J. Shan, K. F. Mak, *Nat. Mater.* **2019**, 18, 324.
[10] R. Karplus, J. M. Luttinger, *Phys. Rev.* **1954**, 95, 1154.
[11] J. Smit, *Physica* **1955**, 21, 877.
[12] L. Berger, *Phys. Rev. B.* **1970**, 2, 4559.
[13] M.-W. Yoo, J. Tornos, A. Sander, L.-F. Lin, N. Mohanta, A. Peralta, D. Sanchez-Manzano, F. Gallego, D. Haskel, J. W. Freeland, D. J. Keavney, Y. Choi, J. Strempfer, X. Wang, M. Cabero, H. B. Vasili, M. Valvidares, G. Sanchez-Santolino, J. M. Gonzalez-Calbet, A. Rivera, C. Leon, S. Rosenkranz, M. Bibes, A. Barthelemy, A. Anane, E. Dagotto, S. Okamoto, S. G. E. te Velthuis, J. Santamaria, J. E. Villegas, *Nat. Commun.* **2021**, 12, 3283.
[14] J. Nichols, X. Gao, S. Lee, T. L. Meyer, J. W. Freeland, V. Lauter, D. Yi, J. Liu, D. Haskel, J. R. Petrie, E.-J. Guo, A. Herklotz, D. Lee, T. Z. Ward, G. Eres, M. R. Fitzsimmons, H. N. Lee, *Nat. Commun.* **2016**, 7, 12321.
[15] A. K. Jaiswal, D. Wang, V. Wollersen, R. Schneider, M. L. Tacon, D. Fuchs, *Adv. Mater.* **2022**, 34, 2109163.
[16] L. Wang, Q. Feng, H. G. Lee, E. K. Ko, Q. Lu, T. W. Noh, *Nano Lett.* **2020**, 20, 2468.
[17] G. Koster, L. Klein, W. Siemons, G. Rijnders, J. S. Dodge, C.-B. Eom, D. H. A. Blank, M. R. Beasley, *Rev. Mod. Phys.* **2012**, 84, 253.
[18] Z. Fang, N. Nagaosa, K. S. Takahashi, A. Asamitsu, R. Mathieu, T. Ogasawara, H. Yamada, M. Kawasaki, Y. Tokura, K. Terakura, *Science* **2003**, 302, 92.
[19] B. Sohn, E. Lee, S. Y. Park, W. Kyung, J. Hwang, J. D. Denlinger, M. Kim, D. Kim, B. Kim, H. Ryu, S. Huh, J. S. Oh, J. K. Jung, D. Oh, Y. Kim, M. Han, T. W. Noh, B.-J. Yang, C. Kim, *Nat. Mater.* **2021**, 20, 1643.
[20] D. Zhang, Y. Wang, N. Lu, X. Sui, Y. Xu, P. Yu, Q.-K. Xue, *Phys. Rev. B.* **2019**, 100, 060403.
[21] K. S. Takahashi, H. Ishizuka, T. Murata, Q. Y. Wang, Y. Tokura, N. Nagaosa, M. Kawasaki, *Sci. Adv.* **2018**, 4, eaar7880.
[22] T. Nan, T. J. Anderson, J. Gibbons, K. Hwang, N. Campbell, H. Zhou, Y. Q. Dong, G. Y. Kim, D. F. Shao, T. R. Paudel, N. Reynolds, X. J. Wang, N. X. Sun, E. Y. Tsymbal, S. Y. Choi, M. S. Rzchowski, Y. B. Kim, D. C. Ralph, C. B. Eom, *Proc. Natl. Acad. Sci. U.S.A* **2019**, 116, 16186.
[23] H. Wang, K.-Y. Meng, P. Zhang, J. T. Hou, J. Finley, J. Han, F. Yang, L. Liu, *Appl. Phys. Lett.* **2019**, 114, 232406.
[24] L. Liu, Q. Qin, W. Lin, C. Li, Q. Xie, S. He, X. Shu, C. Zhou, Z. Lim, J. Yu, W. Lu, M. Li, X. Yan, S. J. Pennycook, J. Chen, *Nature Nanotechnology* **2019**, 14, 939.
[25] B. Vengalis, A. Maneikis, F. Anisimovas, R. Butkut, L. Dapkus, A. Kindurys, *J. Magn. Magn Mater.* **2000**, 211, 35.

[26] J. Dho, N. H. Hur, I. S. Kim, Y. K. Park, *J. Appl. Phys.* **2003**, 94, 7670.
[27] L. M. Wang, J.-H. Lai, J.-I. Wu, Y. K. Kuo, C. L. Chang, *J. Appl. Phys.* **2007**, 102, 023915.
[28] Y. Ying, J. Fan, L. Pi, Z. Qu, W. Wang, B. Hong, S. Tan, Y. Zhang, *Phys. Rev. B.* **2006**, 74, 144433.
[29] F. Tsui, M. C. Smoak, T. K. Nath, C. B. Eom, *Appl. Phys. Lett.* **2000**, 76, 2421.
[30] C. Zeng, Y. Yao, Q. Niu, H. H. Weitering, *Phys. Rev. Lett.* **2006**, 96, 037204.
[31] J. Kötzler, W. Gil, *Phys. Rev. B.* **2005**, 72, 060412.
[32] Y. Tian, L. Ye, X. Jin, *Phys. Rev. Lett.* **2009**, 103, 087206.
[33] D. Hou, G. Su, Y. Tian, X. Jin, S. A. Yang, Q. Niu, *Phys. Rev. Lett.* **2015**, 114, 217203.
[34] A. Asamitsu, Y. Tokura, *Phys. Rev. B.* **1998**, 58, 47.
[35] J. Ye, Y. B. Kim, A. Millis, B. Shraiman, P. Majumdar, Z. Tešanović, *Phys. Rev. Lett.* **1999**, 83, 3737.
[36] G. Shibata, K. Yoshimatsu, E. Sakai, V. R. Singh, V. K. Verma, K. Ishigami, T. Harano, T. Kadono, Y. Takeda, T. Okane, Y. Saitoh, H. Yamagami, A. Sawa, H. Kumigashira, M. Oshima, T. Koide, A. Fujimori, *Phys. Rev. B.* **2014**, 89, 235123.
[37] M. Imada, A. Fujimori, Y. Tokura, *Rev. Mod. Phys.* **1998**, 70, 1039.
[38] K. Binder, A. P. Young, *Rev. Mod. Phys.* **1986**, 58, 801.
[39] J. F. Ding, O. I. Lebedev, S. Turner, Y. F. Tian, W. J. Hu, J. W. Seo, C. Panagopoulos, W. Prellier, G. Van Tendeloo, T. Wu, *Phys. Rev. B.* **2013**, 87, 054428.
[40] S. D. Tiwari, K. P. Rajeev, *Phys. Rev. B.* **2005**, 72, 104433.
[41] D. Nam, K. Jonason, P. Nordblad, N. Khiem, N. Phuc, *Phys. Rev. B.* **1999**, 59, 4189.
[42] T. D. Thanh, D. H. Manh, T. L. Phan, P. T. Phong, L. T. Hung, N. X. Phuc, S. C. Yu, *J. Appl. Phys.* **2014**, 115, 17B504.
[43] E. Jiménez, J. Camarero, J. Sort, J. Nogués, N. Mikuszeit, J. M. García-Martín, A. Hoffmann, B. Dieny, R. Miranda, *Phys. Rev. B.* **2009**, 80, 014415.
[44] S. A. Baily, M. B. Salamon, Y. Kobayashi, K. Asai, *Appl. Phys. Lett.* **2002**, 80, 3138.
[45] S. U. Yuldashev, H. C. Jeon, H. S. Im, T. W. Kang, S. H. Lee, J. K. Furdyna, *Phys. Rev. B.* **2004**, 70, 193203.
[46] R. G. Kepler, R. A. Anderson, *J. Appl. Phys.* **1978**, 49, 1232.
[47] H. Kawamura, *Phys. Rev. Lett.* **2003**, 90, 047202.
[48] G. Tatara, H. Kawamura, *J. Phys. Soc. Japan* **2002**, 71, 2613.
[49] C. D. O'Neill, A. S. Wills, A. D. Huxley, *Phys. Rev. B.* **2019**, 100, 174420.
[50] V. V. Anisimov, J. Zaanen, O. K. Andersen, *Phys. Rev. B.* **1991**, 44, 943.
[51] L. Bellaiche, D. Vanderbilt, *Phys. Rev. B.* **2000**, 61, 7877.
[52] G. Kresse, J. Furthmüller, *Phys. Rev. B.* **1996**, 54, 11169.
[53] J. P. Perdew, K. Burke, M. Ernzerhof, *Phys. Rev. Lett.* **1996**, 77, 3865.
[54] P. Blaha, K. Schwarz, G. K. Madsen, D. Kvasnicka, J. Luitz, *An augmented plane wave+ local orbitals program for calculating crystal properties* **2001**, 60.
[55] S. L. Dudarev, G. A. Botton, S. Y. Savrasov, C. J. Humphreys, A. P. Sutton, *Phys. Rev. B.* **1998**, 57, 1505.
[56] L. Si, W. Xiao, J. Kaufmann, J. M. Tomczak, Y. Lu, Z. Zhong, K. Held, *Phys. Rev. Lett.* **2020**, 124, 166402.
[57] L. Si, O. Janson, G. Li, Z. Zhong, Z. Liao, G. Koster, K. Held, *Phys. Rev. Lett.* **2017**, 119, 026402.
[58] T. Miyake, F. Aryasetiawan, *Phys. Rev. B.* **2008**, 77, 085122.

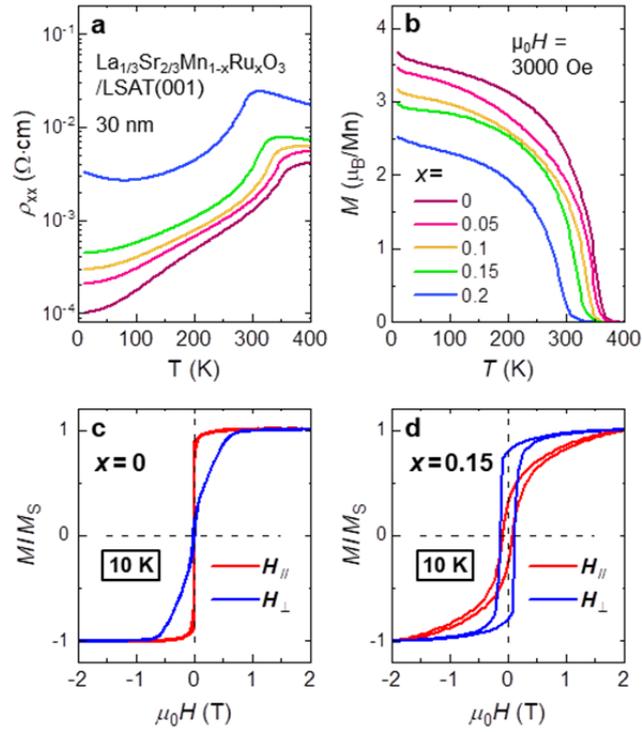

**Figure 1.** Electrical transport and magnetism of LSMRO/LSAT(001) films. a,b) Temperature-dependent resistivity ($\rho$-$T$) (a) and magnetization ($M$-$T$) (b) curves of 30 nm thick LSMRO films with different Ru doping level $x$. c,d) Magnetic field-dependent magnetization ($M$-$H$) curves from the 30 nm thick LSMO and LSMRO ($x$ = 0.15) film at 10 K. The external $H$ is applied along the in-plane and the film normal, denoted as $H_{//}$ and $H_{\perp}$, respectively.

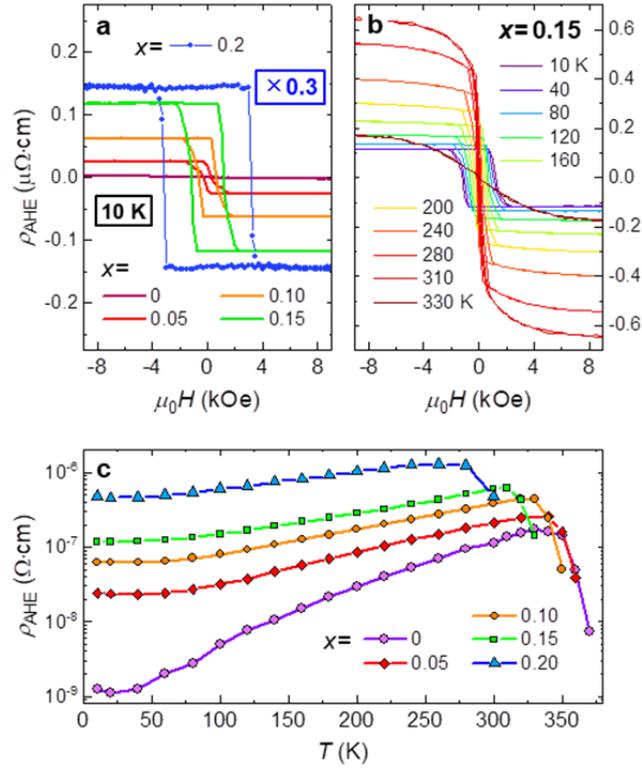

**Figure 2.** Anomalous Hall Effect of the LSMRO films. a) $H$-dependent anomalous Hall resistivity ($\rho_{AHE}$-$H$) curves measured at 10 K from 30 nm LSMRO films with various $x$. $\rho_{AHE}$ values of the LSMRO ($x = 0.20$) film were multiplied by a factor of 0.3 for clarity. b) $\rho_{AHE}$-$H$ curves of the LSMRO ($x = 0.15$) at various $T$. c) $T$-dependent $\rho_{AHE}$ at saturated $H$, extracted from the $\rho_{AHE}$-$H$ curves of 30 nm thick LSMRO films with different $x$.

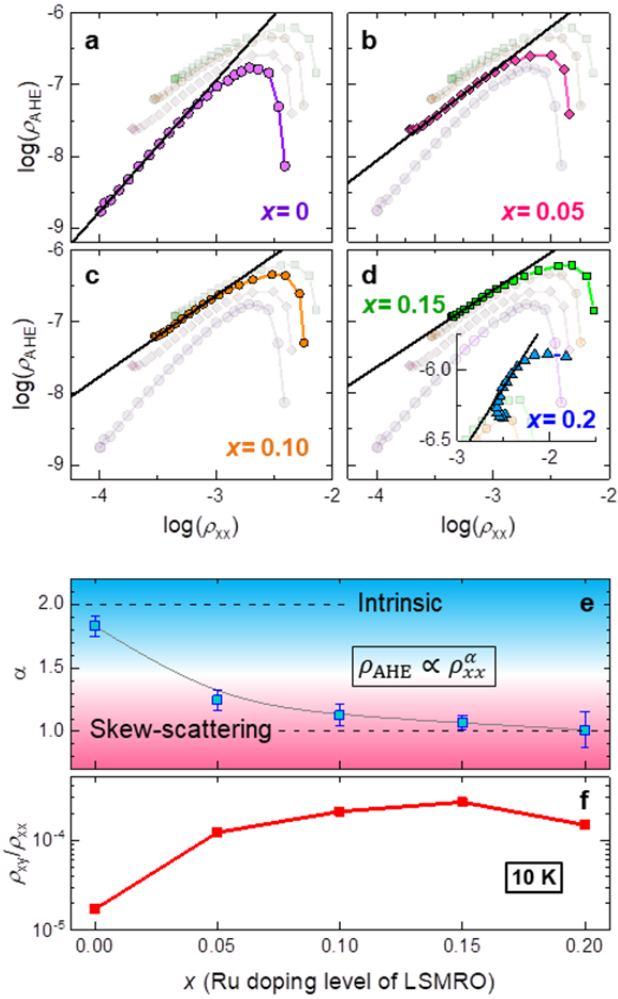

**Figure 3.** Analyses of the mechanism of anomalous Hall effect. a-d) log($\rho_{xx}$)-log($\rho_{AHE}$) curves extracted from $\rho_{AHE}$–$H$ curves of the 30 nm LSMRO films with various $x$. The $x$ = 0.2 curve is inserted in the inset of (d). e) The $x$-dependent $\alpha$ value, which is obtained from the linear fitting in a-d). f) Ru doping level $x$-dependent the anomalous Hall angle $\theta_H$ = $\rho_{AHE}$ / $\rho_{xx}$ at 10 K.

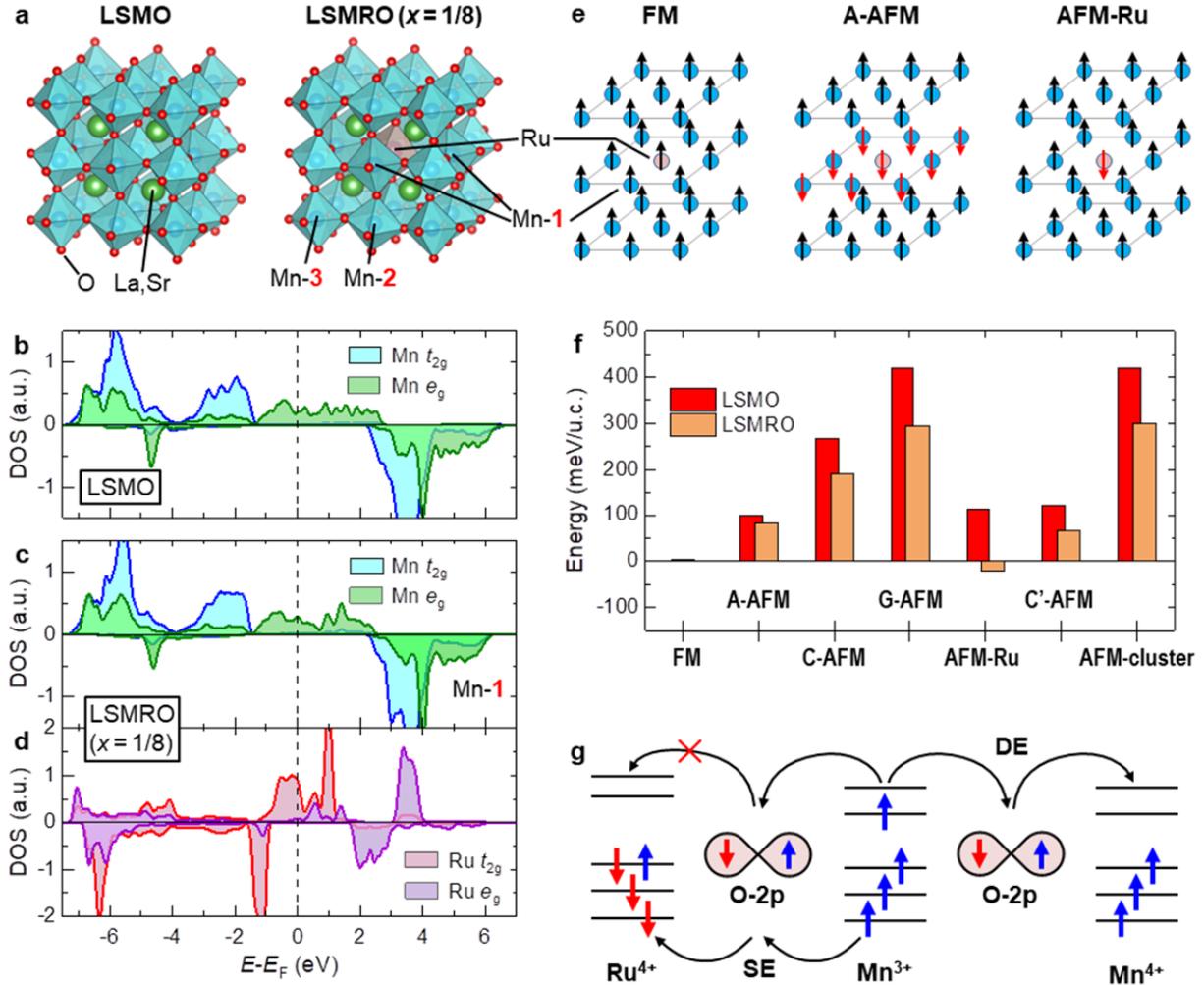

**Figure 4.** First-principles calculations on the electronic and magnetic structures of LSMRO. a) Schematic 2×2×2 supercells of LSMO (left) and LSMRO (right, $x = 1/8$) for first-principles calculations. We added imaginary oxygen atoms outside the supercells to highlight the $MnO_6$ octahedra. By considering the periodic boundary condition, the chemical formula of the supercell is $(La_{2/3}Sr_{1/3})_8Mn_7Ru_1O_{24}$, which corresponds to a 2×2×2 perovskite supercell. b-d) Density of states profiles of the Mn 3d and Ru 4d bands calculated from LSMO (b) and LSMRO (c,d) supercells. e) Possible magnetic structures of the Mn and Ru sites in the LSMRO supercells. f) Energy differences between various spin configurations in the LSMRO supercell. g) Schematic illustrations of the double-exchange (DE) and super-exchange (SE) coupling among $Mn^{3+}$, $Mn^{4+}$, and $Ru^{4+}$ ions.

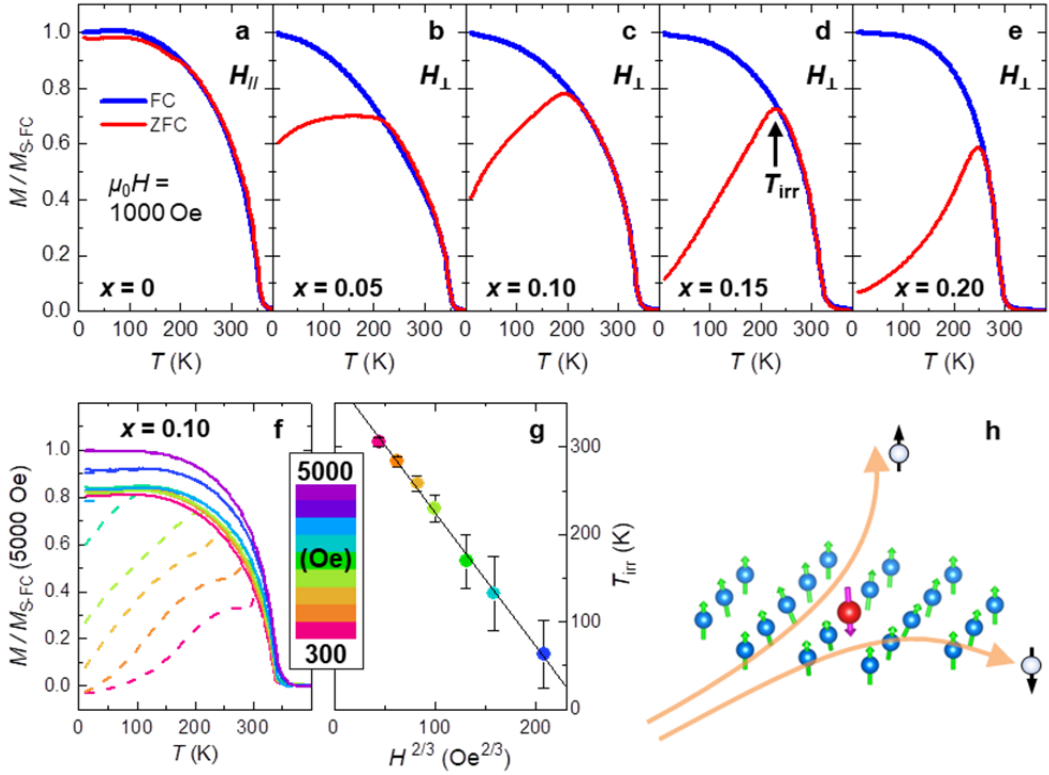

**Figure 5**. Magnetic characterizations of the spin-glass state in LSMRO. a-e) *M-T* curves of 30 nm LSMRO films with various *x*, measured after field-cooling (FC) and zero-field-cooling (ZFC) to 10 K at 1000 Oe. The external *H* is applied along the magnetic easy axes. Namely, *H* is applied in-plane for the LSMO film and along the film normal for all of the LSMRO films. f) FC (solid lines) and ZFC (dashed lines) *M-T* curves of 30 nm LSMRO film (*x* = 0.1) measured under different *H*. g) The irreversibility temperature ($T_{irr}$) plotted as a function of $H^{2/3}$. h) Schematics of the Mn-Ru antiferromagnetic coupling-induced spin frustration and the enhancement of anomalous Hall effect through the skew-scattering mechanism.

# Supporting Information for

# Ru doping induced spin frustration and enhancement of the room-temperature anomalous Hall effect in La$_{2/3}$Sr$_{1/3}$MnO$_3$ films

*Enda Hua, Liang Si, Kunjie Dai, Qing Wang, Huan Ye, Kuan Liu, Jinfeng Zhang, Jingdi Lu, Kai Chen, Feng Jin, Lingfei Wang\*, and Wenbin Wu\**

## Contents:

**Section 1. Structure and magnetism of LSMRO/LSAT(001) films**

    Figures S1-S3

**Section 2. Magnetic anisotropy and anomalous Hall effect in LSMRO/STO(001) films**

    Figures S4-S5

**Section 3. Electronic and magnetic structures of LSMRO films**

    Figures S6-S9, Table S1

**Section 4. Additional results related to spin glass state**

    Figures S10-S12

# Section 1. Structure and magnetism of LSMRO/LSAT(001) films

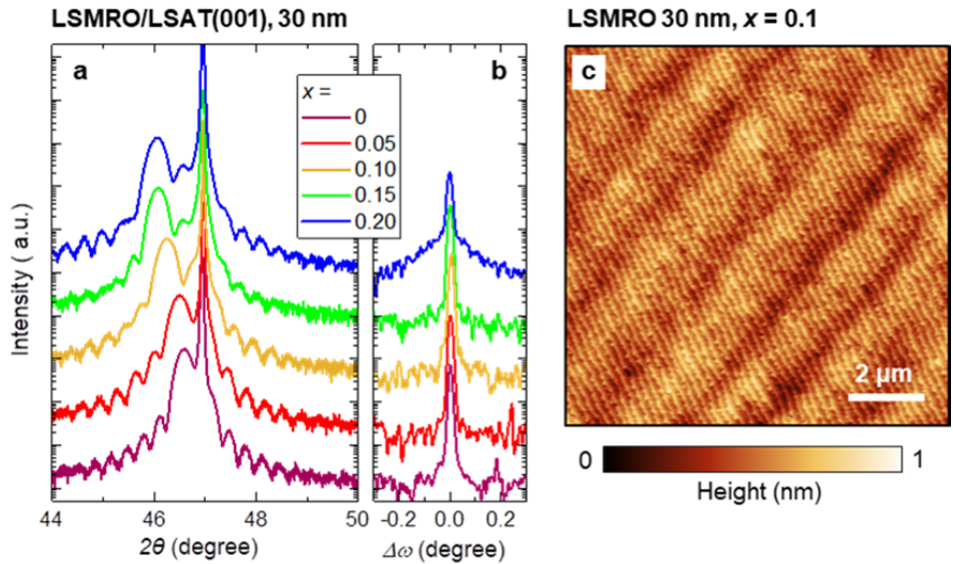

**Figure S1.** Structural characterizations of LSMRO/LSAT(001) films. a,b) XRD *2θ-ω* linear scans (a) and *ω*-scan rocking curves (b) near LSMRO(002) diffractions measured from a set of 30 nm-thick LSMRO/LSAT(001) films with various *x*. The *2θ-ω* linear scans display well-defined Laue fringes around LSMRO(002) diffractions, signifying smooth film surfaces and sharp LSMRO/LSAT interfaces. The full width at half maximum (FWHM) values for all of the rocking curves are below 0.02°, indicating good epitaxial qualities. In addition, the rocking curve for *x* = 0.20 sample shows a broad "shoulder", which may relate to additional disorders and local strain relaxation caused by the high Ru doping concentration. c) Surface topography measured from a representative LSMRO film (30 nm, *x* = 0.1) by atomic force microscopy. The scan area is 10×10 μm$^2$. The smooth surface and well-defined one-unit-cell-high terrace structure further confirm the high epitaxial quality of the LSMRO/STO(001) films.

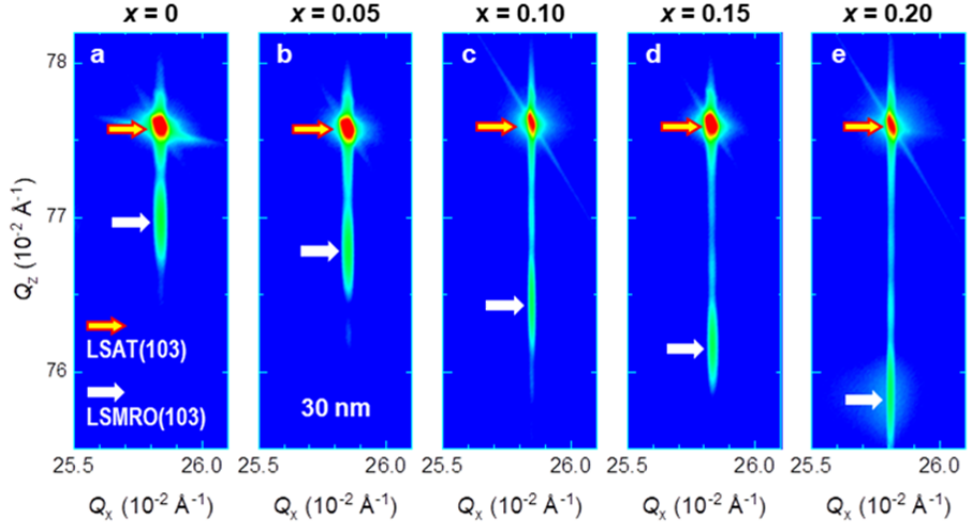

**Figure S2.** Off-specular reciprocal space mappings (RSMs) of the 30 nm LSMRO/LSAT(001) films with various *x*. The film (substrate) diffractions are marked in white (red/yellow) arrows. In all of the RSMs, in-plane reciprocal space vectors ($Q_x$) are the same as those of substrates, demonstrating the coherent strain states. The out-of-plane reciprocal space vectors ($Q_z$) of the film are smaller than that of the substrate. Namely, the out-of-plane lattice constants of the compressively-strained LSMRO films are larger than that of the LSAT(001) substrate. As *x* increases, the gradually increased $Q_z$ further signifies an enhanced in-plane compressive strain.

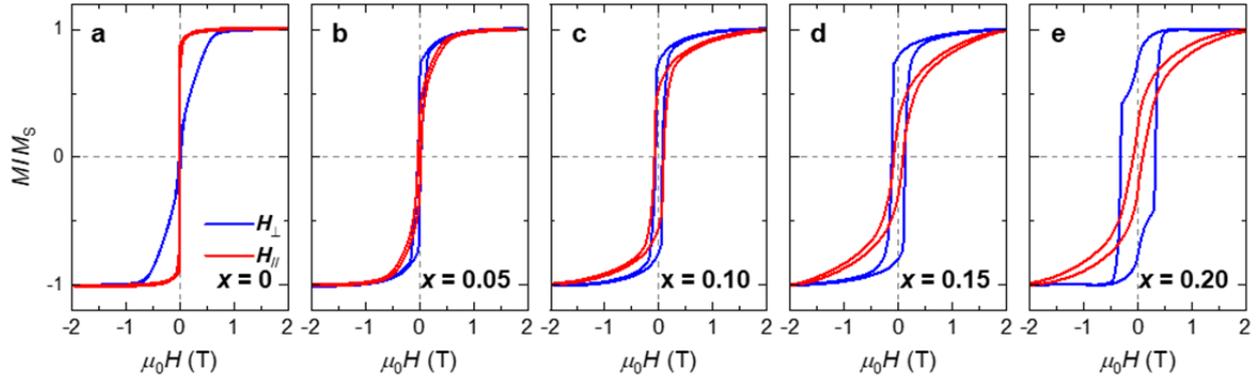

**Figure S3.** Evolution of magnetic anisotropy (MA) in LSMRO films. a-e) Magnetic field-dependent magnetization (*M-H*) curves measured from the 30 nm LSMRO films at 10 K. The external *H* was applied both in-plane or along the film normal (denoted as $H_{//}$ and $H_{\perp}$, respectively). The LSMO film shows an easy-plane MA, while the LSMRO films show a perpendicular MA (PMA). As *x* increases, the saturation field in the *M-$H_{//}$* curves (anisotropic field, $H_A$) and the coercive field in the *M-$H_{\perp}$* curves increases gradually, signifying a gradually enhanced PMA.

**Section 2. Magnetic anisotropy and anomalous Hall effect in LSMRO/STO(001) films**

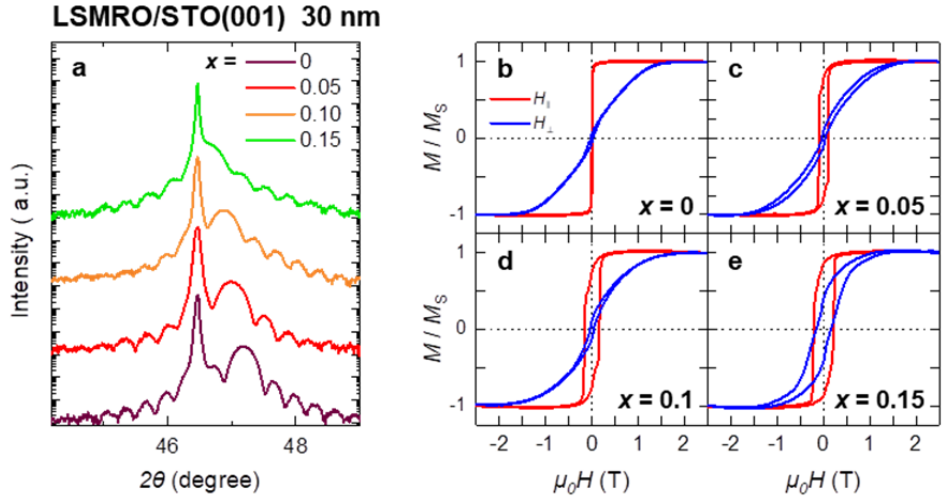

**Figure S4.** Structural and magnetic characterizations of LSMRO/STO(001) films. (a) XRD $2\theta$-$\omega$ linear scans of the 30 nm thick LSMRO/STO(001) films with various $x$. All of the curves display well-defined Laue fringes, signifying a high epitaxial quality. And the LSMRO(002) peaks located at higher Bragg angles, further confirm that the STO(001) substrate can impose a strong biaxial tensile strain on the LSMRO films. (b-e) *M-H* curves measured from 30 nm thick LSMRO/STO(001) films with various $x$ in both $H_\perp$ and $H_{//}$ configurations. The biaxial tensile strain imposed by STO(001) substrates induces strong in-plane MA in the LSMRO films. The anisotropic field ($H_A$) is up to ~1.5 T, in contrast to the PMA observed in the LSMRO/LSAT(001) films.

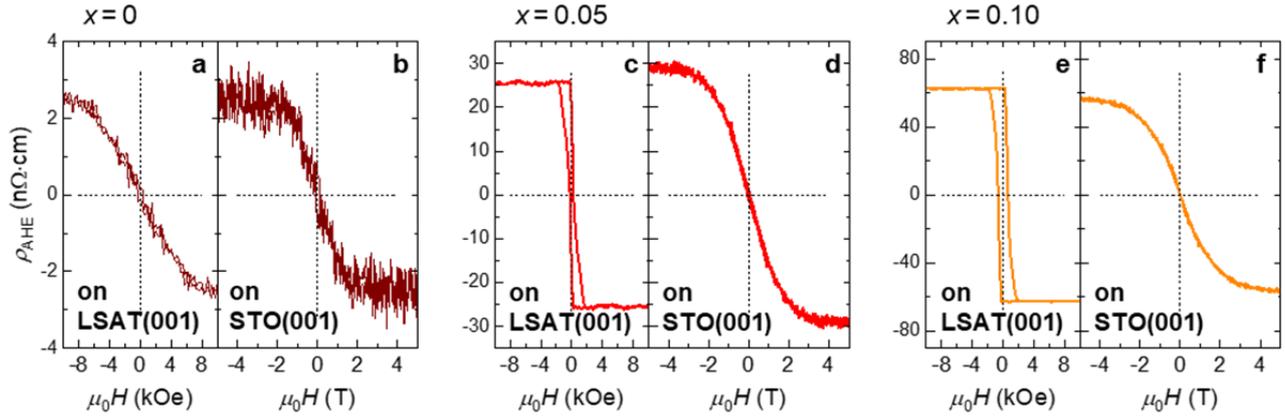

**Figure S5.** Comparison of AHE of LSMRO/LSAT(001) and LSMRO/STO(001) samples. The $\rho_{AHE}$-$H$ curves (at 10 K) of 30 nm LSMRO/LSAT(001) [LSMRO/STO(001)] films with different $x$ are shown in (a), (c), and (e) [(b), (d), and (f)]. Although the curves of LSMRO/STO(001) films become slanted, the saturated $\rho_{AHE}$ values are comparable with the corresponding ones of LSMRO/LSAT(001) films.

## Section 3. Electronic and magnetic structures of LSMRO films

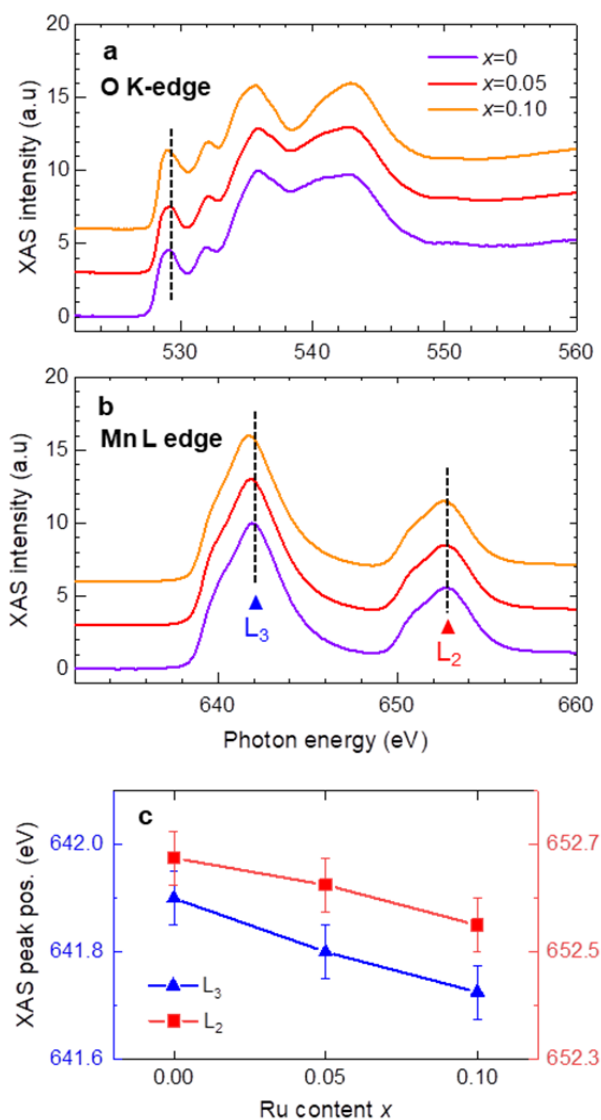

**Figure S6.** X-ray absorption spectroscopy (XAS) of the LSMRO films. a,b) XAS curves near the O-K (a) and Mn-L (b) edges, measured from the 30 nm LSMRO films at room temperature. c) $x$-dependent XAS peak positions extracted from the curves in (b). Both the Mn $L_2$ and $L_3$ peaks exhibit a slight shift towards lower energies as $x$ increases, which indicates that the valence state of Mn is lowered with Ru doping.

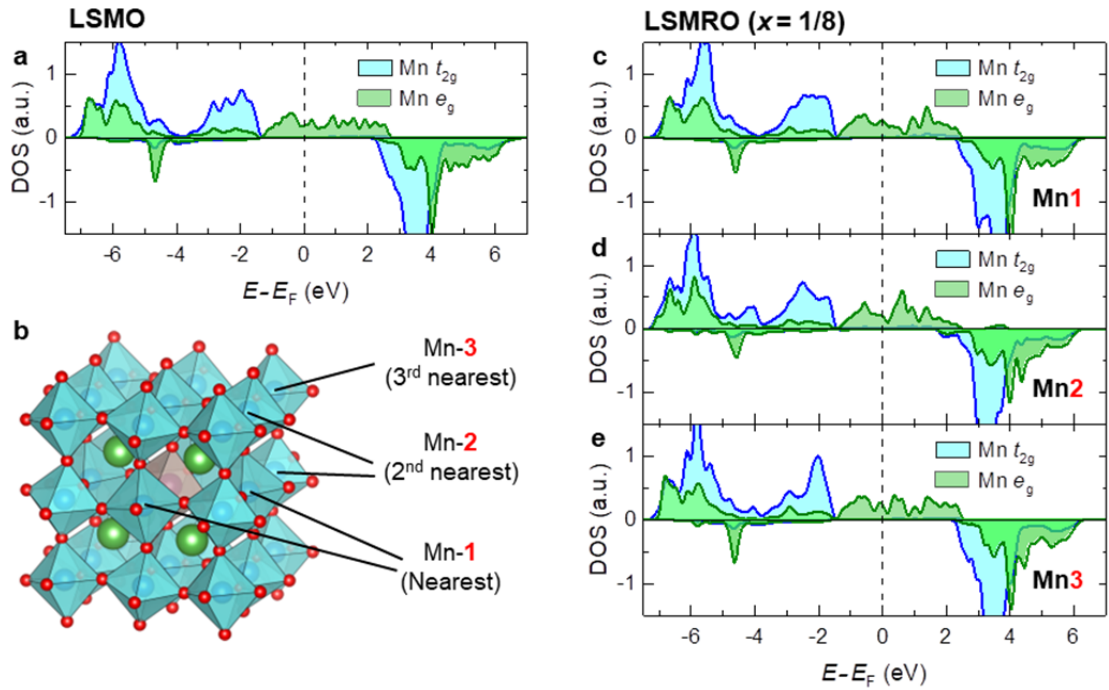

**Figure S7.** First-principles-calculation on the electronic structures of LSMO and LSMRO supercells. a) Spin-polarized density of states (DOS) profiles of the Mn 3$d$ orbitals calculated from the LSMO supercell. b) Schematic supercell structure of the 2×2×2 LSMRO ($x$ = 1/8) supercell. In the LSMRO supercells, the first, second, and third nearest Mn sites to the central Ru are marked as Mn-1, Mn-2, and Mn-3, respectively. c-d) Spin-polarized density of states (DOS) profiles of Mn 3$d$ orbitals at the Mn-1, Mn-2, and Mn-3 sites in the LSMRO supercell.

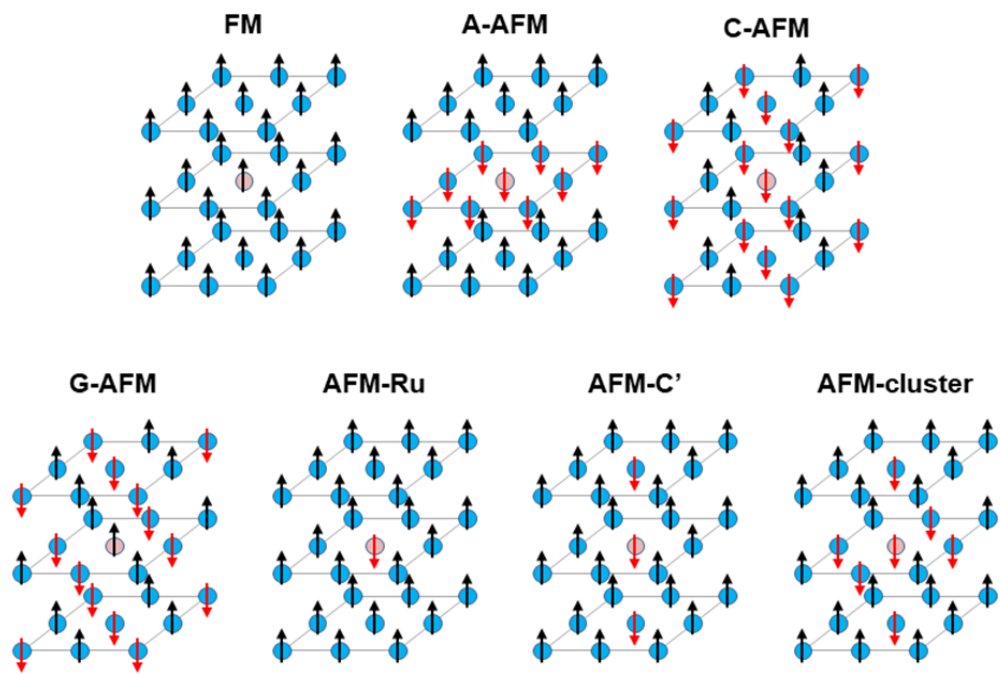

**Figure S8.** Possible magnetic orderings of the LSMRO supercell. Blue (Red) circles represent the Mn (Ru) cations. Black (red) arrows represent the local moments pointing upward (downward). The total energies of these magnetic structures are plotted in Figure 4f in the main text.

| LSMO | | $t_{2g}$ | $e_g$ | total |
|---|---|---|---|---|
| | Mn | 2.074 | 2.313 | 10.849 |
| | | $t_{2g}$ | $e_g$ | total |
| LSMRO | Mn-1 | 2.025 | 2.272 | 10.620 |
| | Mn-2 | 1.997 | 2.190 | 10.373 |
| | Mn-3 | 2.007 | 2.161 | 10.344 |
| | Ru | 3.287 | 3.199 | 16.260 |

**Table S1.** Spread ($\Omega$) of the Wannier function of Mn-*3d* and Ru-*4d* orbitals in LSMO and LSMRO. The unit of $\Omega$ is Å$^2$.

According to previous literature [*Rev. Mod. Phys.* **84**, 1419 (2012)], the spread ($\Omega$) of the Wannier function is an effective parameter to measure the real-space expansion (i.e., the degree of delocalization), of a specific orbital. To compute the $\Omega$ values, we project the Ru-*4d* and Mn-*3d* orbitals (computed by WIEN2K code) near the Fermi level onto maximally-localised Wannier functions (MLWF) using Wannier90 code [*Computer Physics Communications* **178**, 685 (2008)] and WIEN2WANNIER interface [*Computer Physics Communications* **181**, 1888 (2010)]. As summarized in Table S1, we calculated the $\Omega$ values of Mn/Ru $t_{2g}$ and $e_g$ orbitals in the same LSMO and LSMRO supercells for our DFT calculation. The $\Omega$ values of Mn $e_g$ and $t_{2g}$ orbitals in both LSMO and LSMRO supercells are similar and close to 2 Å$^2$, leading to a to total value up to ~10 Å$^2$. By contrast, the $\Omega$ value of both Ru-$t_{2g}$ and $e_g$ orbitals in LSMRO are beyond 3 Å$^2$, leading to a total value of over 16 Å$^2$. Namely, the $\Omega$ of Ru 4*d* orbitals are more than 50% larger than that of the Mn 3*d* orbitals. This result provides solid evidence that demonstrates the Ru 4*d* orbitals are much more delocalized and spatially-expanded than the Mn 3*d* orbitals.

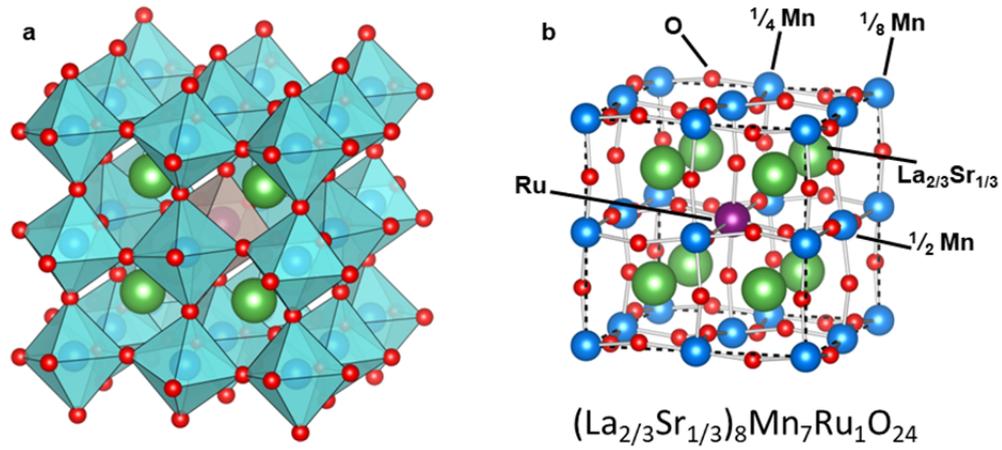

**Figure S9.** Schematic lattice structure of the LSMRO supercell for calculation. (a) Schematic lattice structure reproduced from Figure 4a in Main Text. To highlight the MnO$_6$ octahedra geometry, this version includes some Mn-O bonds outside the supercell. (b) Schematic lattice structure of the real LSMRO supercell for calculation. Although the supercell consists of 9 Mn/Ru atoms, the periodic boundary condition should be considered. The corner-, edge-, and face-shared Mn atoms should be counted as 1/8, 1/4, and 1/2 of a single Mn atom. Therefore, the chemical formula of this LSMRO supercell should be (La$_{2/3}$Sr$_{1/3}$)$_8$Mn$_7$Ru$_1$O$_{24}$, which corresponds to a 2×2×2 perovskite supercell.

**Section 4. Additional results related to spin glass state**

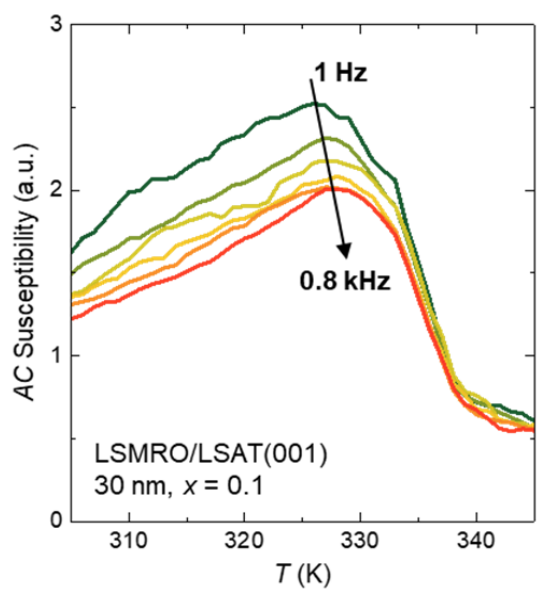

**Figure S10.** *T*-dependent AC susceptibility curves at various frequencies ranging from 1 Hz to 0.8 kHz, measured from the 30 nm LSMRO/LSAT(001) ($x$ = 0.1) film. As the frequency increases, $T_F$ gradually shifts to a higher *T*. This is one of the typical characteristics of the spin glass state.

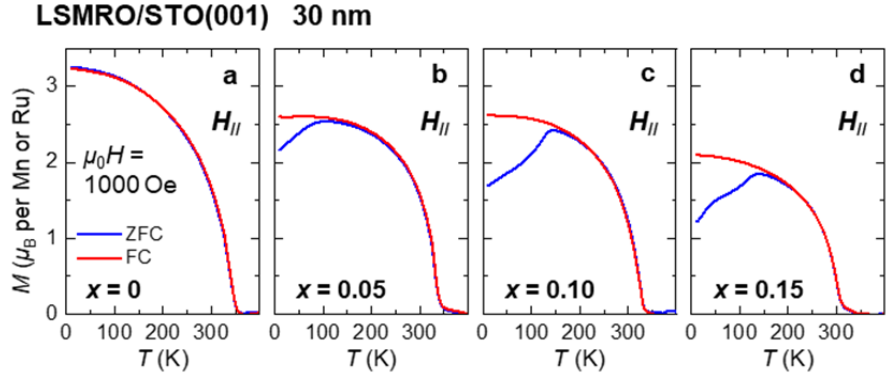

**Figure S11.** *M-T* curves of 30 nm LSMRO/STO(001) films with various *x*, measured after field-cooling (FC) and zero-field-cooling (ZFC) to 10 K at 1000 Oe ($H_{//}$). The FC and ZFC *M-T* curves with *x* > 0 (b-d) show clear bifurcation behaviors, which suggest the existence of a robust spin glass state, similar to the ones observed in LSMRO/LSAT(001) films. These results further highlights the dominating role of Ru-doping in inducing the spin frustration.

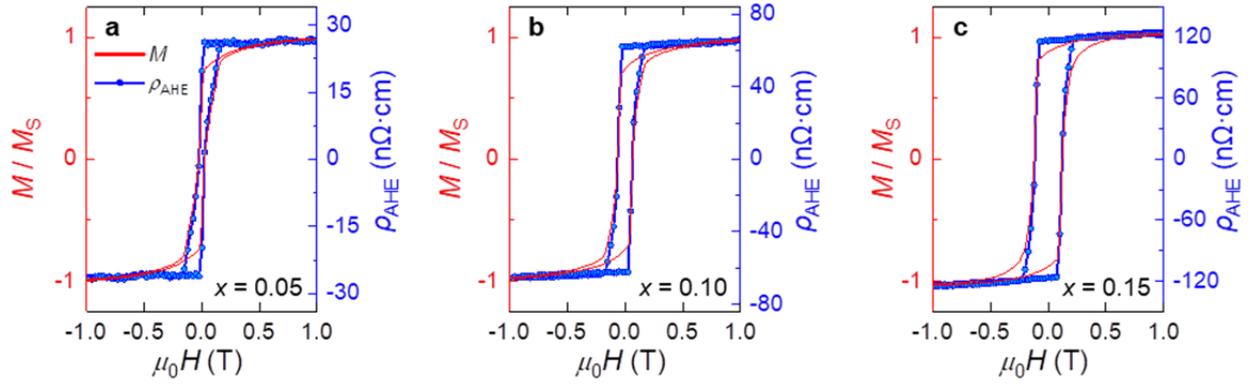

**Figure S12.** Comparisons of the *M-H* and $\rho_{AHE}$-*H* curves of the LSMRO/LSAT(001) films. a-c) Normalized *M-H* (red) and $\rho_{AHE}$-*H* (blue) curves of the LSMRO films with $x$ = 0.05 (a), 0.10 (b), and 0.15(c) at 10 K. The external *H* are applied along the film normal. These two sets of curves coincide very well at high *H* range, while some differences occur near the coercive field ($H_C$). Specifically, the *M-H* curves seem to be more difficult in reaching saturation. We speculate that this detectable inconsistency may originate from the Ru doping-induced spin chirality.